\documentclass[a4paper,10pt]{article}
\usepackage[utf8]{inputenc}
\usepackage{hyperref}
\usepackage[english]{babel}
\usepackage{amssymb,amsmath}

\title{
    Structure of Nucleons and Their Interaction in the Concept of Nonperturbative QCD as a Pressing Issue of 21st-Century Physics
}
\author{
    V.I.~Komarov\footnote{e-mail: komarov@jinr.ru}\\
    Joint Institute for Nuclear Research, Dubna, 141980 Russia
}
\date{}

\begin{document}

\maketitle

\begin{abstract}
The problem of the structure of nucleons and their interaction in the concept of nonperturbative
QCD is discussed as an approach to studying the transformation of current quarks into constituent ones and
the search for the mechanism of such a transformation, creating the bulk of the nucleon mass. Attention is
drawn to the possibilities offered by studying central nucleon-nucleon collisions in this aspect.
\end{abstract}

Modern physics operates with hundreds of elementary
particles. The first parameters of particle identification
are the average value and variance of its mass.
In this case, the mass of nucleons determines the mass
of objects in the world around us. Naturally, the problem
of the origin of mass is the primary fundamental
problem of particle physics. The discovery of the mass
generation mechanism proposed by Braut, Engler,
and Higgs, which was triumphantly confirmed in
experiments, turned out to be the greatest achievement
of particle physics in recent decades. This mechanism
has become a widely known scientific fact. If
you ask how the masses of elementary particles are
generated, most physicists will probably say ``the
Higgs mechanism.'' And that would be wrong. Wrong
because the Higgs boson mechanism generates only
masses of leptons and current quarks and has nothing
to do with the mass of constituent quarks that form
nucleons~\cite{cite1}. This mass is at least 80\% of the mass of
the Universe of the mass of the visible Universe (hadrons,
leptons, and the electromagnetic field), and an
even larger share on Earth.

There is no reliable and generally accepted
approach to describe the generation of nucleon mass.
It is only known that such generation is due to the
spontaneous violation of the chirality of the vacuum,
but the question of the mechanism of the transition
from a ``light,'' 5-MeV, current quark to a ``heavy,''
300-MeV, constituent quark remains completely open.
How the interaction with the vacuum of three high-momentum
current quarks converts their kinetic
energy into the mass of constituent quarks that stick
together into a nucleon is unknown. It is obvious that
the fundamental nature of such a transition is not
inferior to the fundamental nature of the Higgs mechanism.

The discovery of the mechanism of this transition
certainly requires knowledge of the structure of the
nucleon and the dynamics of the interaction of
constituent quarks in the concept nonperturbative
Quantum Chromodynamics (NPQCD). This QCD has
achieved significant success. Effective chiral
Lagrangians made it possible to reproduce all the basic
properties of light mesons and their interactions at
low energies~\cite{cite2,cite3}. The most successful, consistent and
universal microscopic description of light baryonic systems,
achieved in the chiral model of constituent quarks~\cite{cite4},
simultaneously reproduces baryon spectra and
baryon-baryon interactions at low energies. Phase
shifts of $NN$ scattering, binding energies of the lightest
nuclei, wave function and form factors of the deuteron,
masses and channels of strong decay of
dibaryon resonances, cross sections for hyperon-nucleon
interactions, and even certain properties of
pentaquarks~\cite{cite5} are available for this model. However,
NPQCD has no direct derivation from the basic postulates
of QCD, and its achievements do not indicate
a mechanism for the transition of current quarks to
constituent ones.

Information on the structure of nucleons obtained
at high energies in the concept of perturbative QCD is
also not constructive in this respect. Enormous efforts
have been made to date to elucidate the structure of
nucleons in this concept, but the successfully obtained
high-energy structure of nucleons is very far from the
low-energy one. Suffice it to recall that the bulk of the
data is in the form of one-dimensional parton distributions,
while the distributions of quarks in the
ground state of a nucleon are certainly three-dimensional
in space. In recent decades, the theory of generalized
quark distributions, which are a function of
three variables~\cite{cite6}, and attempts are made to relate the
``high-energy'' and ``low-energy'' distributions for a
unified description of the entire energy region (see, for
example~\cite{cite7}).

However, the problem is not only in the dimension
of the distributions. The difference is more fundamental:
High-energy data deal with current quarks, and
low-energy data deal with constituent ones. It would
be possible to obtain information on the distributions
of constituent quarks directly from the distributions of
current ones, only knowing the relationship of current
quarks with their gluon and quark-antiquark environment
in constituent quarks. However, no such information
is available. Therefore, the quark structure of
nucleons in the low-energy region remains a completely
special and self-sufficient problem. Its solution
obviously demands new experimental data, and the
most effective direction is to obtain them in a wide
range of energies, including both low-energy and
high-energy ones.

In this aspect, the NICA complex being created at
JINR has a great advantage, energetically covering
both areas of interest. The transition of the meson-baryon
phase of matter to the quark-gluon phase has
traditionally been studied in collisions of heavy nuclei.
However, for the study of the nonperturbative QCD
structure of nucleons, such collisions create very
significant complications:\\
(1) the low quark density of nuclei in comparison
with the density of nucleons requires high collision
energies to achieve the baryon and energy densities of
interest;\\
(2) fluctuations of the nucleon density in the initial
state of nuclei lead to a significant scatter of local
baryon and energy densities in the intermediate state;\\
(3) a large number of nucleons participating in the
collision creates a complex dynamic scenario of collision,
in particular, with the presence of a mixed
meson-baryon and quark-gluon phase.

Therefore, the study of the elementary process of
transition between constituent and current quarks in
nucleon-nucleon collisions has radically more definite
initial conditions and is more accessible for interpretation
than in collisions of heavy nuclei.

The NICA project provides for the creation of a
facility for the study of nucleon–nucleon collisions
(the Spin Physics Detector, SPD)~\cite{cite8}. One of the priority
goals of this facility may be precisely the study of
the structure of nucleons in the low-energy region,
that is, in the region of nonperturbative QCD, as well
as in the region of the transition to the perturbative
region. Optimal research conditions are created in the
kinematics of collisions, providing overlapping of the
central quark regions of nucleons~\cite{cite9}. Collisions with
large impact parameters determined by meson-baryon
processes are ineffective for studying the quark degrees
of freedom of the arising states. Therefore, it is of
particular importance to distinguish collisions with impact
parameters on the order and less than the radius of the
quark region of nucleons, that is, $r_c \approx 0.4$~fm. The
corresponding criterion for centrality was recently
discussed in~\cite{cite10}.

It should be emphasized that inelastic central collisions
of nucleons have a significant probability, which
does not disappear with increasing energy. R. Feynman
in 1969~\cite{cite11} drew attention to the fact that the
constant with the energy part of the total inelastic
cross section of hadron collisions is determined by
events with a high multiplicity of produced particles.
High multiplicity is now widely used as a criterion for
collision centrality. Therefore, the part of the total
inelastic nucleon collision cross section discussed by
Feynman, which does not vanish with energy, can be
associated with the collision of their quark core.

The total cross section for central nucleon collision
is at the level
$\sigma^\mathrm{inel}_{cNN} = \pi r_c^2 = 5$~mbn.
In the collision of
protons, due to the conservation of the baryon number,
a nucleon pair appears in the final state: $pp$, $np$,
$nn$. Therefore, each collision is accompanied with a
probability close to 100\%, the almost isotropic emission
of at least one proton. As a result, for almost isotropic
emission, the differential cross section for the
emission of one proton at an angle close to 90\textdegree, is
about $\sigma^\mathrm{inel}_{cNN}/4\pi = 0.4$~mbn/sr.

Numerous experimental data show that, in a wide
range of processes, as well as angular and energy
measurement conditions, the emission of protons is
accompanied by the emission of deuterons with a ratio
of about $3 \times 10^{-3}$. If we use the registration of a
deuteron at an angle close to 90\textdegree{}~\cite{cite10}, the process of
interest to us occurs with a differential cross section
$\approx 1 \mu$b/sr, which is quite sufficient for carrying out
correlation and polarization measurements.

Central $NN$ Collisions open up a wide range of
studies on the structure and dynamics of nucleon
interaction in the nonperturbative QCD regime. The
object of research is the characteristics of the reaction
\begin{equation}
p + p \to d(90^\circ) + M,
\label{pp_dM}
\end{equation}
where $M$ is a system of mesons with an admixture of
baryon-antibaryon pairs at sufficiently high energies.

Of primary interest is the search for the effects of
the transition of constituent quarks to current ones. It
is essential that the region of such a transition should
be within the SPD energy range (4--28 GeV). Indeed,
according to long-motivated estimates~\cite{cite12}, the
characteristic impulse of spontaneous chirality breaking
has a value close to $P_\chi \approx 1.2$~GeV/$c$. If the transition
occurs during the destruction of all six colliding constituent
quarks, the total energy required for this in the
QCD is $\sqrt{s_{NN}} \approx 12$~GeV. This means that, in the central
NN collisions in the energy region from the overlapping
energy of the quark content of nucleons,
$\sqrt{s_{NN}} \approx 3$, up to $\approx 12$~GeV, a phase state is created in
which constituent quarks, Goldstone bosons, and gluons
are acting degrees of freedom. This state can be
called a constituent-quark phase of strongly interacting
matter.

The study of the energy dependence of the processes
of central collision of nucleons can be an effective
means of observing the discussed fundamental
transition of constituent quarks to current ones. It can
be expected that such a transition will manifest itself in
the energy dependence of the momentum and multiple
distributions of particles formed in certain channels
of the final state and in their correlations and
polarization characteristics.

Note that, over the past decade, considerable interest
has arisen in the study of central $pp$ collisions at
high energies. However, the energies of hundreds of
GeV~\cite{cite13} and TeV~\cite{cite14} used are too large to observe the
discussed transition of constituent quarks to current
ones, while the unique conditions for such an observation
can probably be provided by SPD.

The study of dibaryon resonances should also be
among the top-priority tasks. QCD spectroscopy
arose in the pioneering work~\cite{cite15} in the 1960s. Light
dibaryon resonances with masses 2.1--2.2 GeV~\cite{cite16,cite17}
were originally viewed as loosely coupled $N\Delta(1238)$
couples. The difficulties that arose in the application
of the meson-baryon description of dibaryon resonances
increased even more after the observation~\cite{cite18}
with respect to heavy resonance with a mass of
2.38 GeV, close to the mass $\Delta(1238)\Delta(1238)$ couples.
This stimulated the successful consideration of
dibaryon resonances in the framework of the chiral
model of constituent quarks~\cite{cite19}. The QCD nature of
the heavier dibaryon resonances is beyond doubt. The
very existence of such resonances with a mass of
3.0~GeV~\cite{cite20} and 2.65~GeV~\cite{cite21} is shown in the experiments
that have already been carried out. A study of
dibaryon resonances in the central mode $NN$ collisions
can become a promising program for the study of
reaction mechanisms determined by the degrees of
freedom of nonperturbative QCD (NPQCD). The
sensitivity of the processes of pion generation in $NN$
collisions in the GeV energy region to such mechanisms
was shown in the works of V.I. Kukulin~\cite{cite22}.
Central collisions can become a test bench for
NPQCD models of the reaction mechanism determined
by the interaction of chiral constituent quarks.

Among the tasks of the first stage, one should also
mention the study of the expected dependence of the
parameters of the produced mesons and meson correlations
on the energy and baryon densities in $NN$
collisions~\cite{cite23,cite24}.

Having limited ourselves to a brief indication of
experiments initiated by modern theoretical expectations,
we should also mention the possibility that completely
new unexpected effects will be manifested, as is
often the case when a new experimental field is
opened.

When discussing the proposed experiments on central
nucleon–nucleon collisions, it should be borne in
mind that the SPD program from the very beginning
of the project was focused on polarization studies of
the nucleon spin structure~\cite{cite25}.

On the whole, the creation of the SPD will open up
opportunities for a broad program of investigating the
structure of nucleons in nonperturbative QCD. It is
difficult to overestimate the possibilities of SPD for
carrying out polarization studies in this aspect. The
accumulation of empirical information and the development
of adequate phenomenological models can
become the basis for creating a rigorous nonperturbative
QCD theory of baryons. Such a program requires,
on the one hand, intensifying the theoretical work in
the discussed problems and, on the other, involving
experimenters in preparing the corresponding experiments
on the SPD. The first step in solving the problem
under discussion should be the realization of its
fundamental nature.

\section*{Acknowledgments}
I thank V.A. Bednyakov, A.V. Kulikov, E.A. Strokovsky,
and D.A. Tsirkov for their interest in the problem and our
helpful discussions. The support provided by V.I. Kukulin,
whose recent passing was untimely, was indispensable.

\end{document}